\documentstyle[twoside,fleqn,espcrc2]{article}

\newcommand{\Dsl}{\Gamma\cdot D}
\newcommand{\half}{{{1\over  2} }}
\newcommand{\vol}{ {\rm vol} }
\newcommand{\CB}{{\cal{B}}}

\newcommand{\CG}{{\cal{G}}}
\newcommand{\CH}{{\cal{H}}}
\newcommand{\CR}{{\cal{R}}}

\newcommand{\CF}{{\cal{F}}}
\newcommand{\CQ}{{\cal{Q}}}
\newcommand{\CL}{{\cal{L}}}
\newcommand{\CM}{{\cal{M}}}
\newcommand{\CN}{{\cal{N}}}
\newcommand{\CO}{{\cal{O}}}

\newcommand{\p}{\partial}
\newcommand{\Tr}{{\rm Tr}}

\newcommand{\IC}{{\bf C}}

\newcommand{\IR}{{\bf R}}
\newcommand{\IZ}{{\bf Z}}
\newcommand{\be}{\begin{equation}}
\newcommand{\ee}{\end{equation}}
\newcommand{\bea}{\begin{eqnarray}}
\newcommand{\eea}{\end{eqnarray}}

\newcommand{\AmS}{{\protect\the\textfont2
  A\kern-.1667em\lower.5ex\hbox{M}\kern-.125emS}}

\title{Integrating over the Coulomb branch in $\CN=2$
gauge theory}

\author{Marcos Mari\~no and Gregory Moore \address{Department of Physics,
        Yale University, \\
        Box 208120, New Haven, CT 06520, USA}%
        \thanks{Supported by
DOE grant DE-FG02-92ER40704.}
}

\begin{document}

\begin{abstract}
We review the relation of certain integrals over the
Coulomb phase of $d=4$, $N=2$ $SO(3)$
supersymmetric Yang-Mills theory with Donaldson-Witten
theory. We describe
a new way to write an important contact term in
the theory and   show
how the integrals generalize to higher rank
gauge groups.

\end{abstract}

\maketitle

\section{Introduction }

In the past 15 years there has been a fruitful,
sometimes dramatic,
math/physics dialogue in the arena of 4-manifold theory
and SYM.
In the early 1980's  Donaldson defined diffeomorphism
invariants of compact, oriented 4-folds $X$ using
Yang-Mills instantons. These invariants are
best organized in terms of a single function
(or formal series),
on the homology of $X$ called the
the Donaldson-Witten  function.  Let
$P\in H_0(X;\IZ)$, $S  \in H_2(X;\IZ)$, then
the Donaldson-Witten function can be written as:
\begin{equation}
Z_{DW}( p P+ v   S  ) = \sum   d_{\ell, r}
{p^\ell \over \ell!} {v^r \over  r!}
\label{genfun}
\end{equation}
In 1988  E. Witten \cite{TFT}
interpreted $Z_{DW}$  as the generating
function of correlation
functions in topological SO(3) SYM:
\begin{equation}
Z_{DW}(v\cdot S + p P) = \biggl\langle \exp[ v\cdot I(S) + p \CO(P)]
\biggr\rangle_{SO(3), \CN=2}
\end{equation}
where $I(S)$ and $\CO(P)$ are certain operators
defined below.
This lead to an interesting reformulation of the
problem of computing Donaldson invariants and,
in 1994,  when the structure of the vacuum
became apparent due to the work of Seiberg and
Witten \cite{seibergwitten},  Witten \cite{monopole} gave a
beautiful and simple expression for Donaldson invariants for
$X$ of $b_{2}^+ >1$ and ``of simple type,'' reproducing
and extending the results of Kronheimer and  Mrowka
\cite{KM}. All known 4-manifolds of $b_2^+>1$ are
of simple type.

With the publication of
\cite{monopole} the  program of deriving
4-manifold invariants from supersymmetric
Yang-Mills theory became a resounding
success. Nevertheless, a few loose ends remained
to be tied up. These chiefly concerned the derivation
of Donaldson polynomials for $4$-manifolds
of $b_{2}^+ =1$. In this case $Z_{DW}$ is not
quite topologically invariant, and several
subtle points must be addressed.
The talk at the Strings 97 conference, entitled ``Donaldson=Coulomb + Higgs,''
and delivered
by one of us reported on
some work done  in collaboration
with E. Witten on the extension of the SYM approach
to the derivation of the Donaldson polynomials for
manifolds of $b_{2}^+=1$. In the course of this
investigation a few new results on 4-manifolds were
obtained. For example, a general formula relating
Donaldson invariants to SW invariants, even for
4-manifolds not necessarily of simple type
was derived (eqs. (35), (36) below).  Using this
one can show that all  4-manifolds of
$b_1=0,b_2^+>1$
are of generalized simple type.

The results reported at Strings 97 have all been
described in detail in \cite{MW}. Some interesting alternative
derivations and new viewpoints on those results have recently
been described by A. Losev, N. Nekrasov and S. Shatashvili
in \cite{LNS}. In this note, after reviewing some
aspects of \cite{MW}, we   make a few
additional   comments on  the ``$u$-plane integrals.''

\section{Topological field theory and the $u$-plane integral }

\subsection{Manifolds of $b_{2}^+ =1$}

Recall that for $X$ compact and oriented the intersection
form on $H^2(X;\IR)$ has
signature $((+1)^{b_{2}^+},  (-1)^{b_{2}^-}) $,
and hence is Lorentzian for $b_{2}^+ =1$.
Examples of such manifolds are
$ S^2 \times S^2, \IC P^2 $ and blowups thereof.
This is also the signature of the operator $*$, and, given
a metric one finds a unique solution up to sign,  of the equation
$ \omega \cdot \omega =1 $.  A choice of
sign corresponds to a choice of orientation of
instanton moduli space \cite{dk}.
Such an $\omega$ is called a period
point. Using the period point we can define selfdual
and anti-selfdual projections of 2-dimensional cohomology
classes: $\lambda_+ \equiv (\lambda,\omega)\omega$,
$\lambda_- \equiv \lambda - \lambda_+$.

\subsection{Donaldson-Witten theory according to Donaldson}
In Donaldson theory as formulated in \cite{don,dk,fm}, one starts with
a principal $SO(3)$ bundle $E \rightarrow X$ over a compact,
oriented, Riemannian four-manifold $X$, with fixed
instanton number $c_2(E)$ and Stiefel-Whitney class $w_2(E)$
($SO(3)$ bundles on a four-manifold are classified up to
isomorphism by these topological data). The moduli space
of ASD connections is then defined as
\be
{\cal M}_{\rm ASD} = \{ A: F_{+}(A)=0 \} /{\cal G},
\label{masd}
\ee
where ${\cal G}$ is the group of gauge transformations. To
construct the Donaldson polynomials, one considers the
universal bundle
\be
{\bf P} = (E \times {\cal A} (E))/({\cal G} \times G),
\label{univbdle}
\ee
which is a $G$-bundle over $({\cal A}(E)/{\cal G}) \times X$ and
as such has a classifying map
\be
\Phi: {\bf P} \rightarrow BG.
\label{classmap}
\ee
 The ``observables" of Donaldson theory are then cohomology
classes in ${\cal A}(E)$ obtained by the slant product pairing
\be
\int_{\gamma} \Phi^{*}(\xi), \,\,\,\ \gamma \in H_{*}(X),
\,\,\,\ \xi \in H^{*}(BG).
\label{obs}
\ee
After restriction to ${\cal M}_{\rm ASD} \subset {\cal A}(E)/{\cal G}$, we
obtain
cohomology classes in ${\cal M}_{\rm ASD}$.
In the case of simply-connected manifolds, we have two
different types of observables
\bea
P \in H_0(X) & \rightarrow & {\cal O}(P) \in H^4({\cal M}_{\rm ASD}),
\nonumber\\
S \in H_2(X) & \rightarrow & I(S) \in H^2({\cal M}_{\rm ASD}).
\label{simplyobs}
\eea
The Donaldson invariants are defined by
\be
d_{l,r}  = \int_{ {\cal M}_{\rm ASD} } ({\cal O}(P))^l \wedge (I(S))^r.
\label{doninvs}
\ee
The main point is that $d_{l,r}$ are metric independent
and hence diffeomorphism invariants of $X$
for $b_2^+>1$.  When $b_2^+=1$ it turns out they
are only piecewise
constant as a function of $\omega$.

\subsection{Donaldson-Witten theory according to Witten}
In \cite{TFT}, Witten constructed a twisted version of $N=2 $  SYM theory
which has a nilpotent BRST charge (modulo gauge transformations)
\be
{\overline \CQ} = \epsilon^{{\dot \alpha} {\dot A}} Q_{{\dot \alpha} {\dot A}}
,
\label{qcharge}
\ee
where $Q_{{\dot \alpha} {\dot A}}$ are the SUSY charges.
Here $\dot \alpha$ is a chiral spinor index and $A$ has
its origin in the $SU(2)$ $\CR$-symmetry.
The field content of the theory is
the standard twisted multiplet:
\bea
&A, \,\ \psi_{\mu} = \psi_{\alpha \dot\alpha}, \,\ \phi; \,\,\,\,\,\,\
D_{\mu \nu}^{+}, \chi_{\mu \nu}^{+} = {\overline \psi}_{\dot \alpha \dot
\beta}; \nonumber\\
&{\overline \phi}, \,\ \eta= {\overline \psi} ^{\dot \alpha}_{ ~\dot \alpha}.
\label{fields}
\eea
where $\half D^+_{\mu\nu} dx^\mu dx^\nu$ is a
self-dual 2-form derived from the auxiliary fields, etc.
All fields are valued in the adjoint representation
of the gauge group.
After twisting, the theory is well defined on any Riemannian
four-manifold, since the fields are naturally interpreted as differential forms
and the ${\overline \CQ}$ charge is a scalar
\cite{TFT}.

The observables of the theory are ${\overline \CQ}$ cohomology classes of
operators, and they can be constructed from zero-form observables
${\cal O}^{(0)}$
using the  descent procedure. This amounts to solving
 the equations
\be
d{\cal O}^{(i)} =\{ {\overline \CQ}, {\cal O}^{(i+1)} \}, \,\,\,\,\ i=0,
\cdots, 3.
\label{descent}
\ee
The integration over $i$-cycles $\gamma^{(i)}$ in $X$ of the operators
${\cal  O}^{(i)}$ is then an observable. These descent equations have a
canonical solution: the one-form valued operator $K_{\alpha \dot\alpha}=
-i \delta_{\alpha}^AQ_{ {\dot \alpha} {\dot A}}/4$ verifies
\be
d=\{ {\overline \CQ}, K\},
\label{keq}
\ee
as a consequence of the supersymmetry algebra. The operators
${\cal O}^{(i)}= K^i {\cal O}^{(0)}$ solve the descent equations
(\ref{descent})
and are canonical representatives.
One of the main results of \cite{TFT} is that the generating functional
(\ref{genfun}) can
be written as a correlation function of the twisted theory,
involving the observables
\bea
{\cal O}(P)&=& { 1 \over 8 \pi^2} {\rm Tr} \phi^2, \nonumber \\
 I(S)&=&  {1 \over 4 \pi^2} \int_{S}
{\rm Tr}[ {1 \over 8} \psi \wedge \psi - {1 \over {\sqrt 2}} \phi F].
\label{tftobs}
\eea
 These operators correspond to the cohomology classes
in (\ref{simplyobs}).  The relation between the above two
formulations of Donaldson-Witten theory is described
in detail in many reviews. See, for examples,
\cite{Witten90,Birmingham,Cordes,Labastida}.

\subsection{Evaluation via low-energy effective field theory}

One way of describing the main result of
the  work of Seiberg and Witten is that the
moduli space of $\bar \CQ$-fixed points
 of the twisted  $SO(3)$ $\CN=2$
theory on a compact 4-fold
has {\it two} branches, which we refer to
as the Coulomb and Seiberg-Witten branches.
On the Coulomb branch the expectation value:
$ { \langle \Tr \phi^2 \rangle \over  16 \pi^2 } = u $
breaks $SO(3) \rightarrow SO(2)$ via the
standard Higgs mechanism. The Coulomb
branch is simply a copy of the complex $u$-plane.
However, at two points, $u=\pm 1$, there is a singularity
where the moduli space meets a second branch,
the Seiberg-Witten branch, which is the
moduli space of solutions of the SW equations
modulo gauge equivalence
$\CM_{SW} = \{ (A^D_\mu, M_\alpha) :
F_+(A^D) = \bar MM , \Dsl M = 0 \}/\CG  $.
Here $A^D_\mu =$ is an $SO(2)$ gauge field
which is a magnetic dual to $A_\mu$.

Because of tunneling,  the partition function
on a compact space is a sum over {\it all} vacuum states.
Hence:
\begin{equation}
Z_{DW} = \bigl \langle e^{ p \CO + v\cdot I(S)} \bigr \rangle
= Z_u+ Z_{\rm SW}\  .
\end{equation}

In \cite{monopole} Witten gave the expression for $Z_{DW}$ in the
case of manifolds of simple type with $b_2^+>1$. In this case,
as we will see in a moment, $Z_u=0$. The simple type
condition means that  the only
non-vanishing Seiberg-Witten invariants are associated to
moduli spaces ${\cal M}_{SW}$ of  dimension zero.  Below a generalization of
Witten's formula will be presented which holds for manifolds
of $b_2^+>0$ and not necessarily of simple type.

\section{Derivation of the $u$-plane integral.  }

We now sketch how to derive the contribution
$Z_u$.  The full details are in
\cite{MW}. The first step involves the
identification of the low energy theory and action.
Then we must map the operators $I(S), \CO(P)$ to
the low energy theory.

The untwisted low energy theory has been described
in detail in \cite{seibergwitten,sdual}. It  is an ${\cal N}=2$
theory characterized by a prepotential $\CF$ which
depends on an ${\cal N}=2$ vector multiplet. The effective
gauge coupling is given by $\tau (a) = \CF''(a)$, where $a$ is
the scalar component of the vector multiplet. The
Euclidean Lagrange density for the $u$-plane theory
can be obtained simply by twisting the physical
theory. It can be written as
\bea
&  {i \over 6 \pi} K^4 \CF (a) + {1 \over 16 \pi} \{ {\overline \CQ},
{\overline \CF}'' \chi (D+ F_+) \}  \nonumber \\
& -  { i {\sqrt 2} \over 32 \pi}
\{ {\overline \CQ}, {\overline \CF}' d * \psi \}  \nonumber \\
&- {{\sqrt 2} i \over 3 \cdot 2^5 \pi} \{ \CQ, {\overline \CF}'''
\chi_{\mu \nu}\chi^{\nu \lambda} \chi_{\lambda}^{~\mu} \}{\sqrt g} d^4 x
\nonumber \\
& + a(u) {\rm Tr} R\wedge R  + b(u) {\rm Tr} R\wedge {\tilde R}
\label{action}
\eea
where $a(u),b(u)$ describe the coupling to gravity, and after integration of
the corresponding
differential forms we obtain terms proportional to the signature $\sigma$ and
Euler characteristic $\chi$ of $X$.

As for the operators, we have
$u = \CO(P)$ by definition.  We may then
obtain the 2-observables from the descent procedure.
The result is that
$I(S) \rightarrow \tilde I (S) = \int_S
K^2 u = \int_S {du \over  da} (D_+  +  F_-  ) + \cdots$
Here $D_+$ is the auxiliary field. It is
important to work with off-shell supersymmetry
because of contact terms. Even though we have the
RG flow $I(S)   \rightarrow \tilde I(S)$ it does not
necessarily follow that
$ I(S_1)I(S_2)  \rightarrow \tilde I(S_1)\tilde I(S_2)$
because there can be contact terms. If $S_1$ and
$S_2$ intersect then in passing to the low energy
theory we integrate out massive modes. This
can induce delta function corrections to the operator
product expansion modifying the mapping to the
low energy theory to
\be
I(S_1)I(S_2)  \rightarrow \tilde I(S_1)\tilde I(S_2)
+\sum_{P\in S_1\cap S_2}\epsilon_P T(P)
\label{ope}
\ee
where $T(P)$ are local operators. Such contact
terms were observed in \cite{wjmp} and were related
to gluino condensates.
The net effect is that the mapping to the low energy theory
is simply:
\bea
&\bigl \langle \exp\biggl[
   p \CO(P)  +  I(S)  \biggr]    \bigr \rangle_{\rm  Coulomb}\nonumber\\
&=
\bigl \langle \exp\biggl[
   2 p u  + \tilde I_{\rm low} (S) + S^2 T(u)
\biggr]    \bigr \rangle_{\rm U(1) }
\label{zcoulomb}
\eea
We will discuss the function $T(u)$ in section 6 below.
The explicit form is not needed for the remaining
derivation of the integral.  When evaluating this
path integral standard arguments involving
$\Delta I = \lambda \{ \CQ, V \} $,
$\lambda \rightarrow \infty$, etc.,  must be applied
with care because $V$ has
monodromy, and we must integrate by parts on the
$u$-plane. Nevertheless, a simple
scaling argument presented in
\cite{MW}  shows that the semiclassical evaluation of
the partition function is exact, so the evaluation of
the RHS of equation (\ref{zcoulomb}) above simply amounts to a
semiclassical evaluation of an $\CN=2$ Maxwell
partition function on a curved manifold.
This boils down to several steps:

$\bullet$ Do the Gaussian integral on $D$.

$\bullet$ Do the Fermion zeromode integral on $\eta,\chi,\psi$.

$\bullet$ Do the photon path integral on $U(1)$ gauge field $A_\mu$.

$\bullet$  Evaluate the coupling of $u$ to the background metric
$g_{\mu\nu}$.

The zeromode
 integral on $a(u)$ finally gives the integral over the
$u$-plane.

All of these steps are relatively straightforward.
A few points which should be noted are:

First, and most importantly,
 since $\eta,\chi$ always appear together and
there is only one $\eta$ zeromode (since it is a scalar)
{\it the Coulomb branch contributes only for
manifolds of $b_2^+=1$.}

A second point is the nature of the photon partition function.
The vev
$\langle \Tr \phi^2 \rangle$ breaks the
$SO(3)$ gauge bundle to a sum of line bundles
$E = (L\oplus L^{-1})^{\otimes 2}
= L^2 \oplus \CO \oplus L^{-2} $
and we must sum over ``line bundles'' $L$:
with $
{ 1\over  2 \pi} F(A)   \rightarrow     2\lambda= c_1(L^2) \in
2 \Gamma\equiv 2 H^2(X;\IZ) +   w_2(E)   $.
Here $w_2(E)$ represents an
't Hooft flux for the SO(3) gauge theory.
The sum over line bundles gives a
Siegel-Narain theta function \cite{sdual,verlinde,avatars}
\bea
& & \int d A_\mu \exp[-S_{\rm Maxwell}]
\nonumber\\
& & \,\,\,\,\,\,\,\,\ = y^{-1/2}
\sum_{\lambda\in \Gamma  }\bar q^{\half \lambda_+^2} q^{- \half \lambda_-^2},
\label{theta}
\eea
where $q=\exp (2 \pi i \tau)$ and $\tau=x+iy$.
There is also a phase factor in the lattice sum
whose origin was explained in \cite{sdual}. It has the
form $\exp [ -i \pi \lambda \cdot w_2(X)]$.  Because of the interactions and
2-observable insertions the sum over
line bundles is proportional to:
\footnote{We ignore some subtle overall phases
described in \cite{MW}.}
\bea
&\Psi= \exp \bigl[ -{1 \over 8 \pi y} ({du \over da})^2 S_-^2 \bigr]
\sum_{\lambda \in \Gamma} (-1)^{\lambda \cdot w_2(X)}
\nonumber\\
& \cdot   \biggl[ (\lambda, \omega) + {i \over 4 \pi y} {du \over da} (S,
\omega) \biggr]
\nonumber\\
&\cdot
\exp \biggl[ -i \pi {\overline \tau} (\lambda_+)^2 - i \pi  \tau (\lambda_-)^2
-i {du \over da} (S, \lambda_-) \biggr]. \nonumber\\
\label{psi}
\eea

A third point is the nature of the coupling to
gravity. This was derived in \cite{sdual}
for the case $N_f=0$ and extended to
theories with matter in \cite{MW}. In order to state the result one must
remember that the $u$-plane integral describes a family of elliptic curves
$E_u$. The coupling to gravity is expressed in
terms of quantities naturally associated to that family:
\be
A(u)^\chi B(u)^\sigma \sim \bigl( {da \over  du} \bigr)^{-\chi/2}
\Delta^{\sigma/8},
\label{gravmeas}
\ee
where $\Delta$ is the discriminant of the elliptic curve
and $da/du$ is a period.
The proof follows \cite{sdual} and is based
on R-charge and holomorphy.

\subsection{The explicit expression for the $u$-plane
integral}

The final result of all the computations of the previous
section is the expression:
\begin{equation}
Z_u(p,S) = \int_{\IC}
{du  d\bar u \over  y^{1/2}}
\mu(\tau) e^{2 p u + S^2 \hat T(u)} \Psi
\label{zuplane}
\end{equation}
where $\mu(\tau)   =    { d \bar \tau \over  d \bar u}
\bigl({d a \over  du } \bigr)^{1- \half \chi  } \Delta^{\sigma/8}$ and
\begin{equation}
{\hat T}(u)= T(u) + { (du/da)^2 \over 8 \pi {\rm Im} \,\  \tau }.
\label{contact}
\end{equation}
Here, for simplicity,
we have assumed that $X$ is simply connected.
The result is extended to nonsimply connected
manifolds in \cite{MW}.

We would like to make a few remarks:

$\bullet$ This expression is also the answer for
generalizations of Donaldson theory obtained by
including hypermultiplets.
In fact, this expression makes sense for
{\it any} family of elliptic curves.

$\bullet$ The expression is not obviously
well-defined because $\tau(u)$ has monodromy.

 $\bullet$ The integrand has {\it singularities} at the
cusps   $u=u_i$ where $\Delta(u_i)=0$ or $u= \infty$.

 $\bullet$ Topological invariance is far from obvious
because of the (A)SD projection: $\lambda = \lambda_+ + \lambda_-$,
$\lambda_+ = (\lambda,\omega) \omega$.

 $\bullet$ The expression for $Z_u$, while impressively complicated,
only depends on the classical cohomology ring of
$X$. It is only part of the answer to $Z_{DW}$.

\subsection{The integral $(21)$ makes sense}

There are two points which must be checked
before we can accept (\ref{zuplane}) as a sensible answer for
$Z_u$. First, the integrand is expressed in
terms of quantities which have monodromy. We must
check that the integrand is in fact well-defined on the
$u$-plane.  Second, the integrand has singularities
at the cusps $u=u_i$, and we must define the integral
carefully by a regularization and limiting procedure.

The first point is easily checked since
$\Psi$ is a modular form of weight
$(\half b_2^+ + 1, \half b_2^-)$.
(The modular invariance of $\hat T(u)$
is crucial for this point.)
For single-valuedness in
theories with fundamental matter it is sufficient
to take  $w_2(E) = w_2(X)$ and if we have adjoint matter
we must have $w_2(X) =0$, i.e., $X$ must be a spin manifold.

The second point is rather delicate.
Near any cusp $u\sim  u_i$ we can
make a duality transformation to
the  local $\tau$-parameter:
$q = e^{2 \pi i \tau}$ such that
$  u \rightarrow u_i $ corresponds to
$q \rightarrow 0$, i.e.,
$ {\rm Im} \tau \rightarrow + \infty$.
If we set $\tau= x + i y$, we can regularize the integral
by introducing a cutoff $\Lambda$ for $y$, and then taking the limit
$\Lambda \rightarrow \infty$ at the end. The behaviour
of the $u$-plane integral is given by:
\bea
 Z_u &\sim&  \lim_{\Lambda \rightarrow +\infty} \int^\Lambda {dy \over y^{1/2}
}
\int_{-1/2}^{+1/2} dx  \nonumber\\
& & \,\,\,\,\,\,\,\,\,\,\,\,\,\,\,\,\,\,\,\,\
 \cdot \sum_{\mu,\nu} q^\mu \bar q^\nu \bigl[1 + \CO(1/y)\bigr].
\label{asymptotic}
\eea
Then, one easily checks that
 $\nu \geq \lambda_+^2 $, so $\lambda_+^2 >0$
gives exponential convergence {\it after} $\int dx$.
Curiously, this corresponds to the regularization
used in evaluating string one-loop diagrams.

\section{Wall-crossing and the relation of Donaldson
and SW invariants.}

\subsection{Metric dependence}

General results in
topological field theory imply that
$T = \{ {\overline \CQ}, \Lambda\}$ and hence
 $\delta Z_u \sim \int d^2u
{\p \over  \p \bar u} (\cdots ) $. However, when
field space is noncompact the total derivative
can be nonzero. This is the case for
Donaldson theory on manifolds of $b_2^+=1$.
Indeed, one may derive the general variational
formula
\bea
& &{d \over  dt} Z_u(\omega(t)) \nonumber\\
 &=&
 \sum_{u_i  } \oint_{u_i} du
   \bigl({d a \over  du } \bigr)^{1- \half \chi  } \Delta^{\sigma/8}
e^{ 2 p u + S^2 T(u)} \Upsilon\nonumber\\
\eea
where an explicit expression for $\Upsilon$ was
derived in \cite{MW}.

Let us define a {\it wall} in the K\"ahler cone
associated to $\lambda\in \Gamma$ by
$\lambda_+  = (\omega,\lambda) =0 $.
For such a metric the line bundle $L$ with
$c_1(L) = \lambda$ admits an
{\it abelian instanton} and the variation of the
integral diverges due to a new bosonic zeromode.
For a fixed correlator of order $\sim p^\ell S^r$
the metric variation of $Z_u$ vanishes, except
when $\omega$ crosses a wall.  Then
$Z_u$ has a discontinuous change:
the integral $Z_u(\omega)$ is
{\it not} topologically invariant.

The essential problem is familiar   from string
theory one-loop amplitudes. It corresponds to
the presence of ``massless singularities.''
Mathematically, at a cusp
$y = {\rm Im}\,\  \tau \rightarrow +\infty$, $q \rightarrow 0$.
one considers the integral:
\bea
& &  \lim_{\Lambda \rightarrow \infty}\int^\Lambda {  dy \over  y^{1/2}}
\int^{+\half}_{-\half} dx \underbrace{ c(d) q^d }_{\rm measure   }
\overbrace{(q \bar q)^{\half \lambda_+^2} q^{ -\half \lambda^2} \lambda_+}^{\rm
Theta\  function}
\nonumber\\
 & & \,\,\,\,\,\,\,\,\ =
{\vert \lambda_+ \vert \over  \lambda_+} {c(d=\half \lambda^2) \over  \sqrt{2}}
\eea
where we have taken  $d = \half \lambda^2$
(otherwise the integral $dx$ gives zero).
The contribution of a cusp $u=u_*$ to the
discontinuity from a wall is thus given in terms of a
residue formula:
\bea
\Delta Z_{u} &\sim &
\oint_{u_*} du
q^{-\lambda^2/2}
 ({da \over  du})^{1-\half \chi}
\Delta^{\sigma/8} \nonumber \\
& & \,\,\,\,\,\,\,\,\ \cdot \exp\bigl[2 p u + S^2 T - i {du \over  da}
(S,\lambda)\bigr]
\label{residue}
\eea

\subsection{Donaldson and SW wall-crossing}

There are two kinds of cusps: $u=\infty$ and
$u=u_i$ on the complex plane. We refer to the
finite cusps as SW cusps.   Correspondingly
there are two kinds of walls:

\begin{eqnarray}
&u=\infty : \,\   \lambda_+ =0, \,\   \lambda\in H^2(X,\IZ) + \half w_2(E)
 \nonumber\\
&u=u_* : \,\  \lambda_+ =0, \,\  \lambda\in H^2(X,\IZ) + \half w_2(X)
\label{wc}\nonumber\\
\end{eqnarray}

The wall-crossing discontinuity of $Z_u$ at $u=\infty$
is:
\bea
\Delta Z_u \sim \Biggl[ q^{-\half \lambda^2} \mu(\tau)
e^{2 p u + S^2 T(u) - i (\lambda,S){du \over  da}  } \Biggr]_{q^0}
\label{gottwc}
\eea
For the special case $N_f=0$,
$u = {\vartheta_2^4 + \vartheta_3^4 \over  2(\vartheta_2 \vartheta_3)^2}$,
$\mu(\tau) = {\vartheta_4^{9-b_-} \over  (\vartheta_2 \vartheta_3)^3}$,
$ {da \over  du } = \half \vartheta_2 \vartheta_3$, and
one easily checks that  this is identical to the
famous formula of  G\"ottsche for the
wall-crossing  of $Z_{DW}$ \cite{gottsche,gz}.

However, at the finite cusps $u=u_*$,
$Z_u$ changes but $Z_{DW}$ {\it does not!}
The change in $Z_u$ is:
\bea
\Delta Z_{u} \sim
\oint_{u_*} {du \over  (u-u_*)} {1 \over  (u-u_*)^{\half d(\lambda)} }
\biggl[1 + \CO(u-u_*) \biggr]
\label{swch}
\eea
where $d(\lambda)  = \lambda^2  - {2 \chi + 3 \sigma \over  4} $
and
$\lambda   \in H^2(X;\IZ) + \half w_2(X) $.

\subsection{Mixing between the branches}

At first equation (\ref{swch})
 might appear to be a problem.
In fact, it fits in quite beautifully with the
general principle: Donaldson = Coulomb + Seiberg-Witten:
\bea
Z_{DW} &=& \bigl \langle e^{ p \CO + I(S)} \bigr \rangle
= Z_{\rm Coulomb} + Z_{SW} \nonumber\\
\label{dch}
\eea
Indeed, note
 that $\lambda   \in H^2(X;\IZ) + \half w_2(X) $ defines
a ${\rm Spin}^c$ structure on $X$ and then
$d(\lambda)  = \lambda^2  - {2 \chi + 3 \sigma \over  4} $
is the dimension of SW moduli space:
$d(\lambda)   = \dim \CM_{SW}(\lambda)$.

The Donaldson polynomials do {\it not} jump at SW walls
so:
\begin{equation}
0 = \delta Z_{DW} = \delta Z_{\rm Coulomb} + \delta Z_{SW}
\end{equation}
The two terms on the RHS are nonvanishing. This is
the mixing of Coulomb and ``Higgs'' branches.

\subsection{Structure of the SW contributions}

The cancellation between the changes of
$Z_u$ and $Z_{SW}$ can in fact be turned to
great advantage to derive the general relation
between the Donaldson and SW invariants
\cite{MW}.

We can compute
$\delta Z_{\rm Coulomb}$ and therefore
find $\delta Z_{\rm Higgs}$. Therefore,    we can learn
about the universal holomorphic functions in
the effective Lagrangian $\CL$ {\it with the monopoles
included}. This Lagrangian comes from a prepotential
${\tilde \CF}_M (a_D)$ for the $N=2$ magnetic
vector multiplet and also includes the coupling to the monopole
hypermultiplet. The full action  has the form:
\bea
&  \{ {\overline \CQ}, W \} +  {i  \over 16 \pi}  {\tilde \tau}_M F \wedge F
+ p(u) {\rm Tr} R\wedge R  \nonumber\\
& \,\,\,\,\,\ + \ell(u) {\rm Tr} R\wedge {\tilde R}
- {i {\sqrt 2} \over 2^7 \cdot \pi } { d {\tilde \tau}_M \over da_D} (\psi
\wedge \psi) \wedge F
\nonumber\\
& + { i \over 3 \cdot 2^{11} \pi} { d^2 {\tilde \tau}_M \over da_D^2} \psi
\wedge \psi \wedge\psi \wedge\psi .
\label{monlagra}
\eea
This is due to the fact that the fourth descendant of the prepotential (which
appears in (\ref{action})) can be written as a ${\overline \CQ}$-exact piece
plus the
terms involving the fields $\psi$, $F$ that we have written in
(\ref{monlagra}). The part of the Lagrangian involving
the monopole hypermultiplet can also be written as a ${\overline \CQ}$-exact
term after
twisting, and is included in $W$. The terms involving $p(u)$, $\ell (u)$ are
again
due to the coupling to gravity.

The action (\ref{monlagra}) describes a TFT of a standard sort and can be
evaluated
using standard localization. The terms involving the fields $\psi$ do not
contribute on simply-connected manifolds, and we will drop them (their effect
has been analyzed in \cite{MW}). We obtain then for the generating functional
of the SW contributions:
\begin{equation}
  Z_{SW}   =\sum_{\lambda \in H^2(X;\IZ) + \half w_2(X)} \sum_{u_*}
\bigl\langle e^{p\CO+I(S)}\bigr\rangle_{\lambda, u_*}
\end{equation}
where
\bea
& & \bigl\langle e^{p\CO+I(S)}\bigr\rangle_{\lambda, u_*} \nonumber\\
&=& \int_{{\cal M}_{\lambda} }
\exp \biggl( 2pu + {i \over 4 \pi} \int_S {d u \over da_D} F
+ S^2 T^{*}(u) \biggr) \nonumber\\
& &\,\,\,\,\  \cdot  {\tilde q}_M ^{-\lambda^2/2} P(u)^{\sigma/8}
L(u)^{\chi/4}.
\label{swcontri}
\eea
In this equation,  ${\tilde q}_M= \exp (2 \pi i {\tilde \tau}_M)$ and, by
definition,
 the SW invariant is:
$SW(\lambda) \equiv
\int_{ \CM_{ \lambda} }(a_D)^{\half d(\lambda) }$.
The integral (\ref{swcontri}) is understood in the sense that
we must expand in $a_D$ and isolate the correct power.

Using the fact \cite{monopole} that
$\Delta SW(\lambda) = \pm 1$ we can now
derive the universal functions ${\tilde q}_M$, $P(u)$, $L(u)$
and see that $T^*(u)=T(u)$.

Now, having obtained these universal functions
{\it we can drop the condition $b_2^+=1$ and give the
Donaldson invariants for all manifolds of $b_2^+>0$}:
\begin{equation}
Z_{DW} = Z_u + \sum_\lambda SW(\lambda) \Xi[\lambda]
\end{equation}
where
\bea
\Xi [\lambda] &=& \sum_{u*} \oint {da \over (a-a_*)^{1 + { 1 \over 2}
d(\lambda) } }
\nonumber\\
& & \,\,\ \cdot \exp[ 2pu-i {du \over da} (S, \lambda) + S^2 T(u)] \nonumber\\
& & \,\,\ \cdot \biggl( {a -a_* \over q}\biggr)^{{1 \over 2} \lambda^2 } ({du
\over da}) ^{\chi/2} \biggl(
{\Delta \over a-a_*} \biggr) ^{\sigma/8}.
\nonumber\\
\label{singucontri}
\eea

This generalizes Witten's famous formula
\cite{monopole}  to manifolds
not necessarily of simple type.
One easily checks that $[\Delta({\p \over   \p 2 p})]^N Z=0$
for $N$ large enough, and
hence {\it all 4-folds are of generalized simple type, for
$b_1=0, b_2^+>1$.}  The result is also interesting for
$X$ of simple type $b_2^+>1$, and $N_f>0$.
Here the answer can be expressed
purely in terms of quantities associated to the
elliptic curve at its points of degeneration:
\bea
Z_{DW} &\sim& \sum_{\lambda, u_*}
SW(\lambda)
  \kappa_*^\delta    ({da \over  du})_*^{-(\delta+\sigma)  }  \nonumber\\
& & \cdot \exp\biggl[ 2 p u_*  +
 S^2 T_*  - i ({du \over  da})_* (S,\lambda)
\biggr]\nonumber\\
\label{nf}
\eea
where
$y^2 = x^3  -    {c_4 \over  48}    x - {c_6 \over  864} $,
$({da \over  du})_*^2     = {c_4(u_*)  \over  2 c_6(u_*)}$,
$\kappa_*  = {c_4^3(u_*) \over  \Delta'(u_*) }$ and $\delta=(\chi+\sigma)/4 $.

\section{Evaluation of the $u$-plane integral and its
basic properties}

The question remains of computability of the
$u$-plane integral. It can be computed in
two ways: Indirectly, using basic properties of
the integral and, at least at $N_f=0$,  directly
using techniques developed in string perturbation
theory.  We first discuss the
indirect method:

  $Z_u$ is clearly very complicated.
However, it is completely determined by four
properties:

1. {\it Wall-crossing:} At order $p^\ell S^r$, $Z_u$  is piecewise polynomial
with known discontinuities (\ref{gottwc}).

2. {\it Vanishing theorems:}
$Z_u(X)  = 0 $ for special $X$'s and
special gauge bundles,
with $\omega$  in special chambers.
An important example is $X= {\bf F_1}$,
the first Hirzebruch surface with
$\omega \cdot f=0$ for the fiber $f$ and
$w_2(E)\cdot f \not=0$.

3. {\it The blowup formula:} This relates the
function $Z_{DW}$ on $X$ to
$Z_{DW}$ on the blowup $Bl_P(X)$.
Combined with 1,2, this gives $Z_u$ on ruled surfaces.

4. {\it Homotopy invariance:} Since $Z_u$ only
depends on the classical ring $H^*(X;\IZ)$ we can
replace $X$ by an algebraic surface.

For simply connected 4-folds of
$b_2^+=1$
 $Z_{DW}$ satisfies exactly the same four properties,
so in those cases
\footnote{For example, if $X$ admits a metric of
positive scalar curvature \cite{monopole}.}  where
 $Z_{SW}=0$ then we can immediately
conclude that the above $u$-plane integral for
$N_f=0$ is an integral representation of the Donaldson
invariants.
\footnote{In fact, this result is logically independent of
any use of physics or path integrals, and is completely
rigorous from a mathematical viewpoint.}

The reason these properties determine the integral
is the following: at least for simply connected manifolds,
we can use homotopy invariance to reduce to the case
that $X$ is a rational surface. Any two rational surfaces,
with any two given metrics, can be related to each other
by blowups, blowdowns, and wall-crossing. Then we can
reduce the computation to the case of $X={\bf F}_1$ in a
chamber where $Z_u=Z_{SW}=Z_{DW}=0$.
We have already discussed wall-crossing. We will
briefly review the vanishing theorems
and blowup formula.

\subsection{Vanishing theorems}

On certain manifolds in special chambers and with
special bundles the integral $Z_{\rm u}$ vanishes.
The intuitive principle behind the vanishing theorems
is simple and physical: It costs a lot of action to confine
nonzero flux in a small 2-cycle.  Consider, for example
a product manifold $b \times f$. For a product metric and
connection the Maxwell action satisfies:
\bea
S = \int_{b \times f} F \wedge * F &=&
 {\vol(b) \over  \vol(f)} \biggl( \int_f F\biggr )^2 \nonumber\\
&+ &
{\vol(f) \over  \vol(b)} \biggl( \int_b F\biggr )^2.
\eea
If the gauge bundle is such that the flux
$\int_f F$ is nonzero for all electric line bundles
(e.g. when  $w_2(E) \cdot f = 1$) then the
action goes to infinity in the limit of small
fibers: $\vol(f) \rightarrow 0$. Hence the
theta function decays exponentially fast:
\begin{equation}
\Theta \sim \sum q^{\half \lambda_+^2}
\bar q^{\half \lambda_-^2} \sim
e^{-y/\epsilon^2} \rightarrow 0.
\end{equation}
Thus the integrand vanishes pointwise.

This principle can be used to
establish vanishing theorems. However, it must
be applied with care since the integration
region is noncompact. There are especially
interesting subtleties at $b_2^-=9$, for
elliptic surfaces. See \cite{MW} and
references therein.

\subsection{The blowup formula}

Roughly speaking, the
 procedure of blowing up a manifold at a smooth point
replaces the point $P$ by a sphere - the exceptional
divisor $B$. It changes the intersection form by
$Q \rightarrow \hat Q = Q \oplus (-1) $.

When the exceptional divisor is small: $\omega \cdot B \sim 0$
it can be replaced by a sum over local operators.
This is quite analogous to the  OPE in
conformal field theory  in which
one replaces a disk (or even a handle) on a surface
by an infinite sum of vertex
operators. In Donaldson theory
the only local BRST invariant operators are in the
ring of polynomials generated by $\CO$ and hence we expect
a formula of the form:
\bea
&\biggl\langle \exp\bigl[ I(S) + t I(B) + p \CO \bigr] \biggr\rangle_{\hat{X} }
\nonumber\\
& = \sum_{k\geq 0}  t^k
\biggl\langle \exp\bigl[ I(S) + p \CO \bigr]   \CB_k(\CO) \biggr\rangle_{  X  }
\label{blowup}
\eea
or more informally,
$\exp[t I(B)] = \sum_{k\geq 0} t^k \CB_k(\CO )$.
In fact, this equation, as well as explicit expressions for
$\CB_k$ can be derived quite straightforwardly from
the $u$-plane integral \cite{MW}. The essential
remark is that in the chamber $B_+=0$ the
$\Psi$-function for $\hat X$ factors as a product
of the $\Psi$-function for $X$ times a holomorphic
function of $u$, which can be interpreted as an
insertion of $0$-observables.

For $N_f=0$ the blowup formulae of
\cite{MW} agree with the results of \cite{fs}
and \cite{gz}.   In \cite{LNS} the blowup
formulae play a central logical role.

\subsection{Direct evaluations}

While the above basic properties indeed determine
the $u$-plane integral completely, they do not lead
to a very effective evaluation of these integrals.
Any correlation function function
$\langle \CO(P)^\ell I(S)^r \rangle_X$ is related to
the chamber $\omega\cdot f =0$ of ${\bf F_1}$ by
a finite number of blowups and wall-crossings.
But the number of walls $\nearrow\infty$ for
$\ell, r \nearrow \infty$.
However, for
 $N_f=0$ a direct evaluation is possible:
$\tau(u)$ maps the $u$-plane to the modular curve
$\CM =  \Gamma^0(4) \backslash \CH$ and one can write the
$u$-plane integral as:
\be
Z_u \sim \int_{\Gamma^0(4)\backslash \CH} {d \tau \wedge d \bar \tau\over  y^2}
 \tilde \mu(\tau)
\exp\biggl\{ 2 p u + S^2
   \hat T(u)  \biggr\} \Psi
\ee
This is related to ``theta lifts'' in number theory or
quantum corrections in
1-loop string amplitudes. The $\Psi$ function is
essentially the Narain theta function for signature $(1, b_-)$,
relevant, for example,
 to compactifications of heterotic
string on $K3 \times S^1$.

The integral can be evaluated directly by   calculations
analogous to those described in, for examples,
\cite{hmi,borcherds96,kontsevich97}.
The explicit answers are given in \cite{MW}.
The case of $X=\IC P^2$ turns out to be
rather amusing. The unfolding technique which is
used in the evaluations of \cite{hmi,borcherds96,kontsevich97}
does not apply to this case. Instead, one must integrate
by parts using a nonholomorphic
modular form  of weight $(3/2,0)$  for $\Gamma_0(4)$
discovered by Zagier:
\bea
& & {\bf G}(\tau,y)   =  \sum_{n\geq 0} \CH(n)  q^n \nonumber\\
& &
+ \sum_{f=-\infty}^\infty q^{-f^2} {1 \over  16 \pi y^{1/2}}
\int_1^\infty e^{- 4 \pi f^2 u y} {du \over  u^{3/2}}
\label{zagier}
\eea
where
\begin{equation}
 \CH(\tau)    = \sum_{n\geq 0} \CH(n) q^n  = -{1 \over  12} + {1\over 3} q^3 +
{1 \over  2} q^4 + q^7 +  \cdots
\end{equation}
is a generating function for Hurwitz class numbers.
In \cite{MW} the $SU(2)$ invariants for $\IC P^2$
were evaluated in terms of $\CH(\tau)$.
Comparing to previous results of G\"ottsche
\cite{gottsche}
leads to an interesting formula for class numbers:
\bea
& &
\sum_{n\geq 0} \biggl( \CH(4 n) +  \half \CH(16n)   \biggr)q^{2n} \nonumber\\
&+ & \sum_{n\geq 0}    \half \CH(16n+8)   q^{2n+1}
\nonumber\\
&-&\sum_{n\geq0} \biggl( \CH(4 n+3)
+  \half \CH(16n+12)          \biggr) q^{(4n+3)/2}
\nonumber\\
&-&  \sum_{n\geq 0}     \half \CH(16n+4)  q^{(4n+1)/2}
 \nonumber\\
&= &\sum_{n_1>0,n_2\geq n_1}  (-1)^{n_1 + n_2} (2n_2+1)n_1 \nonumber \\
& & \cdot
{q^{\half(n_2(n_2+1)-n_1^2 )+ 1/8}
\over  \eta^3 }  - {\vartheta_2^4 + \vartheta_3^4 \over  8 \vartheta_4}
\label{classnumbers}
\end{eqnarray}

\section{Remarks on the contact term $T(u)$}

We now return to the contact term $T(u)$. By working
with off-shell supersymmetry in (\ref{ope}), $T(u)$
it is guaranteed to
be ${\overline \CQ}$-closed and is hence locally a
holomorphic function of
$u$.
In \cite{MW} $T(u)$ was  determined
by the requirement that the function ${\hat T}(u)$ given in
(\ref{contact}) is invariant under the
${\rm SL} (2, {\bf Z})$ duality
group and by some asymptotic constraints.
It was rederived in \cite{LNS}
for a larger class of observables from a different point of
view, but the argument is only simple for massless theories
and $N_f<4$.  Here we present yet a third derivation.
We restrict attention to the contact terms for 2-observables
arising from the quadratic Casimir, but the method applies
to arbitrary gauge group including matter with arbitrary
masses.
To find $T(u)$, one first notices that the prepotential of
$SU(2)$, $N=2$ supersymmetric gauge theories verifies
\be
{\partial \CF \over \partial \tau_0}= { 1 \over 4} u,
\label{matone}
\ee
where in the asymptotically free
theories $\tau_0$ is defined by
$\Lambda_{N_f}^{4-N_f} = {\rm e}^{ i \pi \tau_0}$ and
$\tau_0$ is the microscopic coupling for the $N_f=4$
theory (the first definition is of course motivated
by the RG equation).  The relation (\ref{matone}) has been
derived in many different contexts \cite{matone,ey,sty,dp,amz}
and holds for any matter content and bare masses
for the hypermultiplets. We will denote
derivatives of the prepotential with respect to the
variables $a,\tau_0$ by the correspondig subindices.
The duality transformation
\be
\gamma = \left(\begin{array}{cc} a & b \\ c & d
 \end{array}\right) \in {\rm SL}(2, {\bf Z} )
\label{sltrans}
\ee
shifts the second term in (\ref{contact})  by
\be
-{ 4 i \over \pi}
{{ c {\cal F}_{a \tau_0}^2} \over c \tau + d}.
\label{shift}
\ee
But this is precisely the structure of the shift for
${\cal F}_{\tau_0 \tau_0}$:
\be
{\cal F}_{\tau_0 \tau_0} \rightarrow {\cal F}_{\tau_0 \tau_0}-
{{ c {\cal F}_{a \tau_0}^2} \over c \tau + d}.
\label{shifttwo}
\ee
as one can check using the duality transformation
properties of the prepotential or following the approach in \cite{amz}.
We then see that the contact term is given by
\be
T(u)= {4 \over \pi i} {\partial ^2 \CF \over \partial \tau_0^2}.
\label{contactterm}
\ee
An explicit expression for all the cases $0 \le N_f \le 4$ can be
obtained by using the
form of the elliptic curves and the Seiberg-Witten
abelian differentials  (for the
asymptotically free theories with arbitrary masses, the
expression has been obtained in \cite{amz} in a different
context). It is given by
\be
T(u) =-{1 \over 24} E_2 (\tau) \bigl( { du \over da } \bigr)^2+ {1 \over 3}
\bigl( u + \delta_{N_f,3}{ \Lambda_3^2 \over 64} \bigr)
\label{af}
\ee
in the case of the asymptotically free theories, and
\be
T(u)= -{1 \over 24} E_2 (\tau) \bigl( { du \over da } \bigr)^2 +
E_2 (\tau_0) {u \over 3} + {1 \over 9} R E_4 (\tau_0)
\label{fourfla}
\ee
in the $N_f=4$ case, where $R = \sum_{a} m_a^2/2$ and
$E_2$, $E_4$ are the normalized Eisenstein series.
These expressions are
valid for the theories with arbitrary hypermultiplet masses.
The same procedure can be applied to the higher rank
theories for the contact term coming from the
second quadratic Casimir, again with any hypermultiplet
content and arbitrary masses.

Finally, we would like to notice that the parameter $\tau_0$
naturally arises in the context of Whitham hierarchies
as a slow time variable. This should provide a link between
this approach to the contact terms and the one in \cite{LNS}.

\section{Extension to higher rank and other $u$-plane
integrals}

The $u$-plane integral can be also analyzed in the
case of higher rank gauge groups \cite{mm} as a tool to
analyze the higher rank analogues of Donaldson
invariants.  We will consider for simplicity the case
of $SU(N)$, although the formalism can be easily extended
to other compact Lie groups.  The integral is given by
\bea
& & Z(p,S;m_i,\tau_0) \nonumber\\
&=& \int_{\CM_{\rm Coulomb}}
[da  d\bar   a]
A(u)^\chi B(u)^\sigma  {\rm e}^{U + S^2  T_V} \Psi.
\nonumber\\
\label{hrintegral}
\eea
In this equation, $U=\sum_{I=2}^r p^I u_{I}$ is a linear combination
of  the Casimirs
of the group. The 2-observable is
derived from the quadratic Casimir $V=u_2$  and $T_V$ is the
corresponding contact term, given by
equation $(49)$ above.
A simple generalization of the argument
of \cite{sdual}  using holomorphy, modular
properties and R-charge fixes the $A$, $B$
functions to be the natural generalizations
of  (\ref{gravmeas}):
\be
A^\chi=\alpha^\chi \biggl( \det {\p u_I \over  \p a^J} \biggr)^{\chi/2},
\,\,\,\
B^\sigma = \beta^\sigma  \Delta_\Lambda^{\sigma/8},
\label{hrfactors}
\ee
where $\Delta_{\Lambda}$ is the quantum discriminant
associated to the genus $r$ hyperelliptic curve
and $\alpha,\beta$ are constants
on $\CM_{\rm Coulomb}$. (Equation (\ref{hrfactors})
was independently derived in \cite{LNS}.) The lattice
sum $\Psi$ is given in this case by the finite-dimensional
integral
\bea
&\Psi =  \sum_{\lambda \in \Gamma}
 \int \prod_{I=1}^r d\eta^I d \chi^I \int_{-\infty}^{+\infty} \prod_I
d b^I  \nonumber\\
&\exp\biggl[ - i \pi \bar \tau_{IJ} (\lambda_+^I , \lambda_+^J)  -
i \pi \tau_{IJ} (\lambda_-^I , \lambda_-^J)
\nonumber\\
&+  {1 \over  8 \pi} b^I ({\rm Im} \tau)_{IJ} b^J - i V_I (S, \lambda_-^I) -{i
\over 4\pi} V_I
(S, \omega) b^I \nonumber\\
& -  { i \sqrt{2} \over  16 \pi} {\overline \CF}_{IJK} \eta^I \chi^J (b^K + 4
\pi
\lambda_+^K)  - i \pi ({\vec \lambda} \cdot {\vec \rho}, w_2(X))\biggr].
\nonumber\\
\label{hrpsi}
\eea
Here, $\chi$, $\eta$ are Grassmann variables and  $b^I$ are commuting
variables (they have their origin in the zero modes of the $\chi$, $\eta$
fields, and in the auxiliary fields, respectively). In the lattice $\Gamma$ we
have to consider
non-abelian magnetic fluxes (or generalized Stiefel-Whitney classes) that are
associated to the conjugacy classes of $\Lambda_{\rm weight} /\Lambda_{\rm
root}$ \cite{vw}. We sum then over
root vectors of the form
\be
{\vec \lambda}= \lambda^I \vec \alpha_I, \,\,\,\ \lambda^I=
\lambda^I_{ Z} + (C^{-1})^I_{~~J } \pi^J \ ,
\label{hrvectors}
\ee
where $\lambda^I_{Z}$, $\pi^I \in H^2(X, {\bf Z})$,  $\vec \alpha_I$,
$I=1, \cdots, r$, is a basis of simple roots, and $C$ is the Cartan matrix.
The weight $\vec \rho$
appearing in (\ref{hrpsi}) is half the sum of the positive roots,  and the term
involving it is
 the appropriate
generalization of the phase derived in \cite{sdual} for the rank one case (this
term has been derived independently in \cite{LNS}).

Although the higher rank integral (\ref{hrintegral}) is quite complicated, it
can be
analyzed in some detail.
 One can check single-valuedness of the integrand
under  the quantum monodromy group. The proper
definition of the integral
is rather subtle because of the nature of the
singular loci of the moduli space.
The superconformal loci are especially subtle.
One can derive  wall-crossing formulae, which are
generically integrals of a residue. There is
``wall-crossing for wall-crossing''  arising from the
contributions of  codimension two  submanifolds,
and so forth. Using this one can generalize
the above result for $Z_{DW}$ to higher rank
gauge groups. A detailed presentation of these
remarks will appear in \cite{mm}.

\section{Conclusion: Future directions}

There are many interesting future directions in
this subject. We mention just two here.
First, it appears that an  analogue
of $Z_u$ can be written for {\it any} special K\"ahler
geometry. The question remains as to the physical
significance. Can we always find a physical system which
is computing some invariants through $Z_u$?
Examples of systems which could be especially
interesting include the effective theory on the D3
probe in F-theory and the integrable system
associated to variation of Hodge structures
introduced by Donagi and Markman
\cite{donagi}.

Another direction, involving applications
to  Gromov-Witten theory, has recently been proposed in
\cite{LNS}.
%
%

\section{Acknowledgements}

GM  would like to thank E. Witten for the extremely
fruitful collaboration which lead to the results of
\cite{MW} and for many important remarks related to
the new results presented above.  He also thanks
the organizers of the Strings 97 meeting for
the opportunity to speak and for their patience in
procuring a manuscript.

\end{document}